# Memory System Designed for Multiply-Accumulate (MAC) Engine Based on Stochastic Computing

Xinyue Zhang, Yuan Wang*, Yawen Zhang, Jiahao Song, Zuodong Zhang, Kaili Cheng, Runsheng Wang* and Ru Huang
Institute of Microelectronics and Key Laboratory of Microelectronics Devices and Circuits (MoE)
Peking University, Beijing 100871, P.R. China
*Email: wangyuan@pku.edu.cn , r.wang@pku.edu.cn

*Abstract*—Convolutional neural network (CNN) achieves excellent performance on fascinating tasks such as image recognition and natural language processing at the cost of high power consumption. Stochastic computing (SC) is an attractive paradigm implemented in low power applications which performs arithmetic operations with simple logic and low hardware cost. However, conventional memory structure designed and optimized for binary computing leads to extra data conversion costs, which significantly decreases the energy efficiency. Therefore, a new memory system designed for SC-based multiply-accumulate (MAC) engine applied in CNN which is compatible with conventional memory system is proposed in this paper. As a result, the overall energy consumption of our new computing structure is 0.91pJ, which is reduced by 82.1% compared with the conventional structure, and the energy efficiency achieves 164.8 TOPS/W.

## I. Introduction

Convolutional neural networks (CNN) are frequently used in computer vision and natural language processing in recent years [1]. State-of-the-art performances are achieved on various tasks through CNN solutions but at the cost of high computational complexity and power consumption [2]. Due to strict demands on hardware resources and power consumption of edge applications, a new computing paradigm is urgently required.

Stochastic computing (SC) [3] is a promising candidate which realizes arithmetic operations with simple logic gates. For example, multiplication can be performed with a single AND gate. Hence SC has been successfully applied to computation-intensive applications such as neural networks. Multiply-accumulate (MAC) is the basic operation of CNN but consumes large power, while SC can do multiplications and additions with simple logic gates thus becomes an excellent choice for low power CNN applications. Besides, stochastic numbers have better error tolerance compared with binary numbers, since each bit of the stochastic number has a uniform weight [4], and it functions well even under aggressive voltage scaling [5]. However, the conversion between binary and stochastic number consumes huge power, which weakens the advantage of SC.

On the other hand, conventional memory system is optimized and designed for binary computing, consequently it demands extra conversion between stochastic and binary numbers when applied to stochastic modules. Ref. [6] proposed a new memory system specifically designed for SC called StochMem. However, StochMem stores data with analog memory [7], which usually costs a large energy and is not compatible with the conventional memory system. Besides, analog memory has an accuracy problem because of the divergence between read and written values. Therefore, a new memory structure is proposed for SC-based MAC engine in this paper, which greatly decreases the power consumption and is compatible with binary memory system.

The rest of the paper is organized as follow: Section II introduces the basic concept of SC, conventional structure of SC-based MAC engine and motivation of this work; Section III describes the proposed new memory structure and details of each module. Section IV shows the simulation and comparison results; Finally, conclusions are drawn in section V.

## II. Background

*A. Stochastic Computing (SC)*

SC technique processes data in form of bitstreams, where the probability of observing 1 in the bitstream is treated as the value of the stochastic number. For example, bitstream A=01011100 contains four 1s and four 0s, which means the value of A is 4/8. The main advantage of SC is that arithmetic operations can be implemented with simple logic gates [4]. For example, multiplication can be realized with a single AND gate as shown in Fig.1(a). If P1 is the probability of observing 1 in bitstream A, and P2 for B, when A and B are inputs of an AND gate, it is obvious that the probability of seeing 1 at the output is P1×P2. Similarly, addition can be implemented with a MUX gate. Therefore, SC is successfully applied to computation-intensive applications such as digital signal processing [8], artificial neural networks (ANN) [9] and decoding of modern error-correcting codes [10]. Another inherent feature of SC is high error tolerance [4], a single bit-flip in a long bitstream only causes a small change of value and different flips tend to cancel each other out [11]. For example, a bit-flip in A=01101010 (4/8) turns the value into 3/8 or 5/8, each circumstance causes only 1/8 deviation. But a bit-flip leads to large variation in binary numbers if it happens to the most significant bit.

However, conversion between binary and stochastic numbers consumes huge power and easily weakens the inherent benefits of SC. Conventional binary to stochastic converter (BSC) composes of a linear feedback shifting register (LFSR) and a comparator (Fig.1(b)), and stochastic to binary converter (SBC) is always realized by a counter (Fig.1(c)). Both of them take up a large proportion in the overall power consumption. Therefore, works on low power converters [12] and new computing

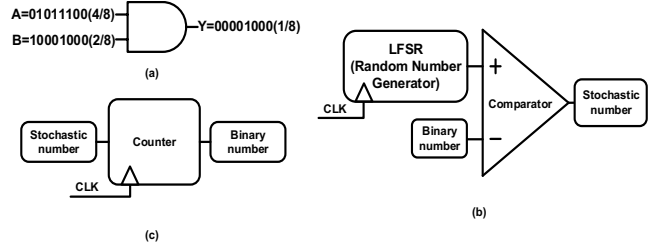

Figure 1. (a) SC multiplication, (b) the structure of conventional BSC, (c) the structure of conventional SBC.

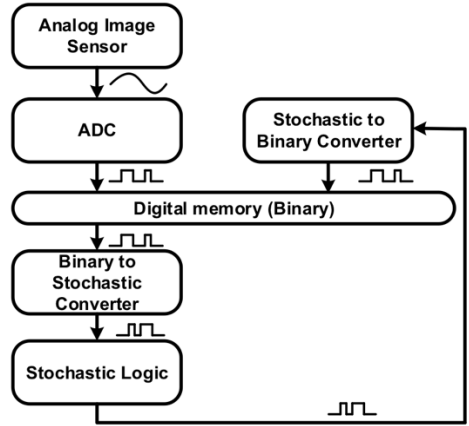

Figure 2. Conventional memory structure of SC [6].

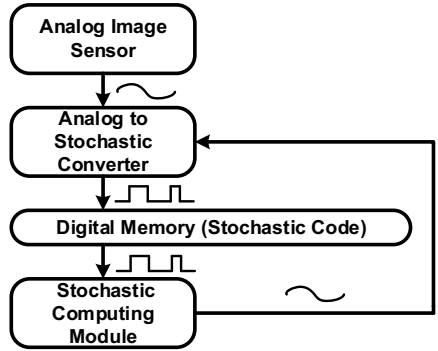

Figure 3. Proposed memory structure designed for SC-based MAC.

structure with simple conversion procedures are of great importance.

*B. Conventional Structure of SC-based MAC Engine*

Fig.2. shows the conventional structure of SC memory system [6]. Input data from analog image sensor is converted into binary numbers through analog to digital converter (ADC) and then stored in standard SRAM array. Then BSC is employed to convert binary numbers into stochastic bitstreams, and next comes to the stochastic logic module, which composes of AND gates array (to do multiplications) and multiplexers (to sum the products up). SC additions are commonly realized by approximate parallel counter (APC) or MUX array, and we choose MUX to do accumulation here as a comparison because it costs relatively low energy [13]. Finally, the output stochastic numbers run into SBC to prepare for next cycle's calculation.

*C. Motivation of This Work*

MAC is an essential but energy-intensive part in CNN, while SC technique can simplify complex computation units to simple logic gates. Therefore, a SC-based MAC engine is a promising implementation.

However, it can be concluded from the conventional structure that, although stochastic logic requires simple logic and is low-cost on its own, assistant modules such as ADC, BSC and SBC do consume a lot of power. According to the statistics, data conversion between binary and stochastic numbers consumes 80% of power and hardware cost of the system [14]. So in the next section, we propose an energy efficient memory structure specifically designed for SC-based MAC engine which greatly decreases the cost of data conversion overhead.

### III. PROPOSED MEMORY STRUCTURE DESIGNED FOR SC-BASED MAC ENGINE

*A. New Memory System Structure*

Conventional memory system shown in Fig.2 is designed and optimized for binary computing and thus demands extra data conversion modules. Ref. [6] proposed an analog memory system called StochMem, which is designed for SC to reduce extra conversion cost. However, analog memory is hard to design, and is incompatible with digital memory. What is worse, the result is not accurate sometimes because of the discrepancy between read and written values. Therefore, we propose a new structure specially designed for MAC operations which stores stochastic numbers. Both of binary and stochastic numbers are digital signals so that the new memory system is compatible with the conventional one. Fig.3 shows the new structure, in which the data from sensor is converted into stochastic numbers with an analog-to-stochastic converter (ASC) and stored in digital memory, then a specially designed MAC stochastic module takes stochastic numbers as input and outputs analog MAC result. The pleasant outcome is that the mixed-signal MAC module is much more energy efficient than conventional stochastic logic.

*B. Mixed-signal MAC Stochastic Module*

Ref. [15] shows a capacitor array which performs MAC operation in binary neural network (BNN). To support high accuracy CNN applications, we modify this structure into a SC-based MAC module with multi-bit inputs, weights and analog outputs. IN and W are both m-bit stochastic numbers (W has an extra sign bit), $IN_i[j]$ ($W_i[j]$) indicates the j-th bit of the i-th IN(W), and $SIGN_i$ indicates the sign bit of $W_i$. N pairs of IN and W do multiplications and then N products are accumulated. Fig.4 shows the mixed-signal MAC stochastic module. EN is an enable signal which turns to high when doing calculations. The mixed-signal structure does multiplication with AND gates and accumulation with a capacitor array. First, when S1 is on and S2 is off, the capacitor array performs accumulation operations of positive and negative products separately through voltage dividing,

$$VP = \frac{n_p}{m \times N + 1} V_{DD}$$

$$VN = \frac{m \times N - n_n}{m \times N + 1} V_{DD}$$

$(n_p = \sum_{i=1}^{N} \sum_{j=1}^{m} IN_i[j] \times W_i[j]$ (when $SIGN_i$=1),
$n_n = \sum_{i=1}^{N} \sum_{j=1}^{m} IN_i[j] \times W_i[j]$ (when $SIGN_i$=0)).

Then, when S2 is on and S1 is off, charge is shared between the 2 tail capacitors,

$$VP = VN = \frac{1}{2} \left\{ \frac{m \times N + (n_p - n_n)}{m \times N + 1} \right\} V_{DD}.$$

A signed MAC operation is performed through this structure. Parameter N and m can be determined according to the accuracy requirement. What's more, the mixed-signal MAC stochastic module consumes less power than conventional SC logic, which will be shown in Section IV.

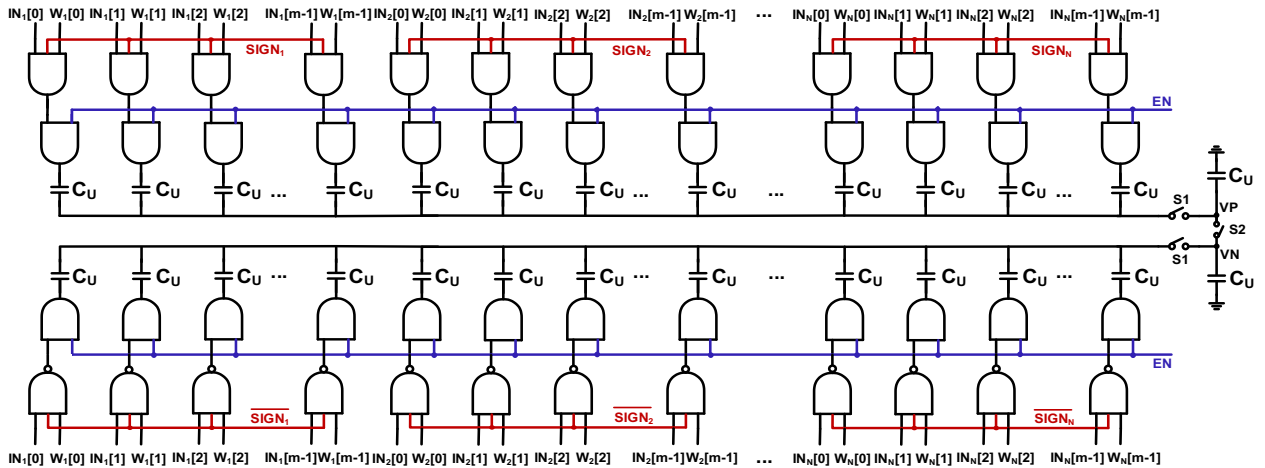

Figure 4. Structure of the proposed mixed-signal SC-based MAC stochastic module.

*C. Analog to Stochastic Converter(ASC)*

There are several ways of transforming analog data into stochastic numbers. ASC presented in [16] performs the conversion with MTJ, which consumes less power than BSC with the same precision but is incompatible with standard CMOS technology. Thus an energy efficient CMOS ASC is proposed here. It is proved that randomness is not necessary for SC and what really matters is the expression method of probability [17]. And deterministic patterns such as thermometer coding achieve better accuracy (even completely accurate) with shorter bitstream length. Fig.5(a) shows the proposed ASC with thermometer coding pattern (3-bit as example), $VREF_i$ is obtained through capacitor voltage divider. When the voltage of $VREF_i$ is higher than analog input, the output of the corresponding sense amplifier (SA) Y[i] will be low. It can be conclude from the coding pattern shown in Fig.5(b) that, if Y[0] equals to 0, Y[1] and Y[2] will also equal to 0, thus a MUX is applied to output Y[1] and Y[2] directly and $SA_1$ and $SA_2$ will be disabled to reduce power consumption. The result of convolution layer has a peak distribution around 0 in CNN applications, thus the proposed ASC reduces energy consumption by 17.9% according our simulation results.

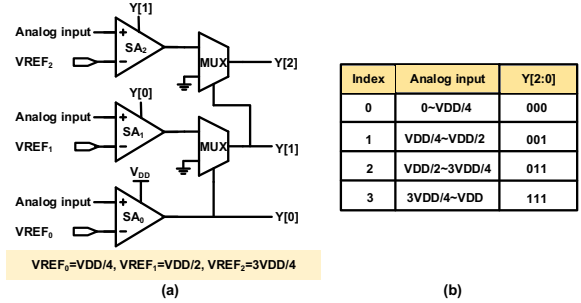

Figure 5.  (a) Structure of the proposed ASC, (b) the pattern of thermometer code.

IV. SIMULATION AND COMPARISON RESULTS

For comparison, the conventional structure and the new structure are both simulated under 28nm technology with the same operation frequency (10MHz).

Fig.6 shows the comparison between the conventional and the proposed memory structures, which illuminates that the proposed structure saves 2 conversion modules (ADC and SDC), and the other 2 modules consume less power than the conventional counterparts. In terms of the memory module, the two memory structure demand the same size of the SRAM array in BNN (0/1) and ternary neural networks (TNN) (0/+1/-1). In multi-bit CNN applications, although the size of stochastic number SRAM array is $(2^n - 1)/n$ times of the conventional one (n denotes the input precision), our structure is still more energy efficient because the energy consumption of ADC also increases with the precision in an exponential way. The simulation result shows that the overall energy consumption is decreased by 82.1% compared with conventional structure. What is more, the mixed-signal MAC stochastic module does not suffer from the low accuracy of MUX in the conventional structure and the accuracy loss of the analog memory in StochMem.

Simulation results of a 15-bit, 300-input system are shown in Table.1. Overall, the proposed MAC engine consumes 0.91pJ and the energy efficiency of the proposed structure achieves 164.8 TOPS/W. The FoM indicates the energy consumption per output quantization steps normalized by the number of elementary arithmetic operation [19], and it reaches 0.38 fJ/step.

| Conventional structure (@10MHz) | | This work (@10MHz) | |
|---|---|---|---|
| Module | Energy | Module | Energy |
| SRAM cell[18] | 28.00 fJ | SRAM cell[18] | 28.00 fJ |
| ADC(4-bit) | 2.15 pJ | \ | \ |
| SBC (4-bit LFSR+4-bit Comparator) | 141.61 fJ | \ | \ |
| BSC (4-bit counter) | 185.54 fJ | ASC(15-bit) | 16.20 fJ |
| SC logic (15-bit) | 20.26 fJ | Mixed-signal SC module (15-bit) | 11.86 fJ |

Figure 6. Energy comparison of each module.

TABLE I. SIMULATION RESULTS OF THE PROPOSED STRUCTURE

|  | This work |
|---|---|
| Technology | 28 nm |
| Application | CNN |
| Computing Paradigm | Stochastic Computing |
| Output Rate | 10 MHz |
| Supply | 1.0 V |
| Power | 9.1 μW |
| Efficiency | 164.8 TOPS/W |
| FoM | 0.38 fJ/Step |

## V. CONCLUSION

SC is a promising computing paradigm applied to CNN for its simple logic of realizing arithmetic calculation and good error tolerance. A new memory system structure is proposed in this article to decrease the extra data conversion costs of SC. As the simulation results show, the proposed structure reduces energy consumption by 82.1%, and is compatible with the conventional digital memory. The simulated energy efficiency achieves 164.8 TOPS/W, which is beneficial to low power applications.

ACKNOWLEDGMENT

This work was supported by National Natural Science Foundation of China (Grant No.61834001 and No.61421005) and the 111 Project (B18001). The authors would like to thank Weikang Qian for the helpful discussions.